\documentclass[reprint,superscriptaddress,amsmath,amssymb,]{revtex4-2}
\usepackage{CJK}
\usepackage{graphicx}
\usepackage{dcolumn}
\usepackage{bm}
\usepackage{amsmath}
\usepackage{amssymb}
\usepackage{xcolor}
\usepackage[colorlinks,
            linkcolor=red,
            anchorcolor=blue,
            urlcolor=red,
            citecolor=blue
            ]{hyperref}
\usepackage[marginal]{footmisc}

\begin{document}
\begin{CJK*}{UTF8}{gbsn}

\title{Anomalous second magnetization peak in 12442-type RbCa$_2$Fe$_4$As$_4$F$_2$ superconductors}

\author{Xiaolei Yi(易晓磊)}
\affiliation{School of Physics, Southeast University, Nanjing 211189, China}
\affiliation{College of Physics and Electronic Engineering, Xinyang Normal University, Xinyang 464000, China}
\author{Xiangzhuo Xing(邢相灼)}
\email{xzxing@qfnu.edu.cn}
\affiliation{School of Physics, Southeast University, Nanjing 211189, China}
\affiliation{School of Physics and Physical Engineering, Qufu Normal University, Qufu 273165, China}
\author{Yan Meng(孟炎)}
\affiliation{School of Physics, Southeast University, Nanjing 211189, China}
\affiliation{School of Physical Science and Intelligent Engineering, Jining
University, Qufu 273155, China}
\author{Nan Zhou(周楠)}
\affiliation{School of Physics, Southeast University, Nanjing 211189, China}
\affiliation{Key Laboratory of Materials Physics, Institute of Solid State
Physics, HFIPS, Chinese Academy of Sciences, Hefei, 230031, China}
\author{Chunlei Wang(王春雷)}
\affiliation{College of Physics and Electronic Engineering, Xinyang Normal
University, Xinyang 464000, China}
\author{Yue Sun(孙悦)}
\email{sunyue@seu.edu.cn}
\affiliation{School of Physics, Southeast University, Nanjing 211189, China}
\author{Zhixiang Shi(施智祥)}
\email{zxshi@seu.edu.cn}
\affiliation{School of Physics, Southeast University, Nanjing 211189, China}


\begin{abstract}
The second magnetization peak (SMP) appears in most superconductors, and is
crucial for the understanding of vortex physics as well as the application.
Although it is well known that the SMP is related to the type and quantity of
disorder/defects, the mechanism has not been universally understood.
In this work, we selected three stoichiometric superconducting
RbCa$_2$Fe$_4$As$_4$F$_2$ single crystals with identical superconducting
critical temperature $T_\textup{c}$ $\sim$ 31 K and similar self-field critical current density $J_\textup{c}$,
but with different amounts of disorder/defects, to study the SMP effect.
It is found that only the sample S2 with moderate disorder/defects shows
significant SMP effect. The evolution of the normalized pinning force density
$f_\textup{p}$ demonstrates that the dominant pinning mechanism changes from the weak
pinning at low temperatures to strong pinning at high temperatures.
The microstructure study for sample S2 reveals some expanded Ca$_2$F$_2$ layers
and dislocation defects in RbFe$_2$As$_2$ layers.
The normalized magnetic relaxation results indicate that the SMP is strongly
associated with the elastic to plastic (E-P) vortex transition.
As temperature increases, the SMP gradually evolves into a step-like shape and then becomes a sharp peak
near the irreversibility field similar to what is usually observed in low-temperature superconductors.
Our findings connect the low field SMP of high-temperature superconductors and the high field peak of low temperature
superconductors, revealing the possible universal origin related to the E-P
phase transition.

\end{abstract}


\maketitle
\end{CJK*}

\section{Introduction}
Research of vortex dynamics in superconductors is of great significance for
fundamental research as well as for technical applications \cite{1857, 428}. One
of the most interesting phenomena of vortex dynamics is the second magnetization
peak (SMP) effect in the magnetization hysteresis loop that has been widely
observed in many type-II superconductors, particularly in high-temperature
superconductors. Until now, various
theoretical models have been proposed to understand the SMP effect, including
softening of vortex lattice \cite{2025, 2026}, vortex lattice structure phase
transition \cite{1910, 342, 346}, a crossover from elastic to plastic (E-P) vortex
transition \cite{1905, 347}, and vortex order-disorder transition \cite{1909},
etc. However, the mechanism is still unclear, and there is still an ongoing
research in recently discovered superconductors \cite{1280, 1880, 1815, 1839}.

For most high-temperature superconductors, the parent compounds are
non-superconducting, and superconductivity is generally achieved by extra element
doping \cite{1600, 1637}. As a consequence, equivalent or nonequivalent chemical
doping inevitably introduces defects and inhomogeneities that can be regarded as
scattering centers for quasiparticles \cite{2127}. It is known that the nature of
vortex is highly sensitive to the type and quantity of defect. In general, the
vortex motion in high-temperature superconductors is determined by the strong
pinning attributed to sparse nanometer-sized defects \cite{1916} and the weak
collective pinning by atomic-scaled defects \cite{1857, 366}. Therefore, all the
inhomogeneities, defects, and scattering centers will act together as the pinning
sources, complicating the behavior of the vortices. In this sense, the stoichiometric
superconducting materials provide an unique platform to study the effect of disorder/defect on the vortex dynamics, free from the additional extrinsic effects
introduced by chemical substitutions.

\begin{figure*}
\includegraphics[width=40pc]{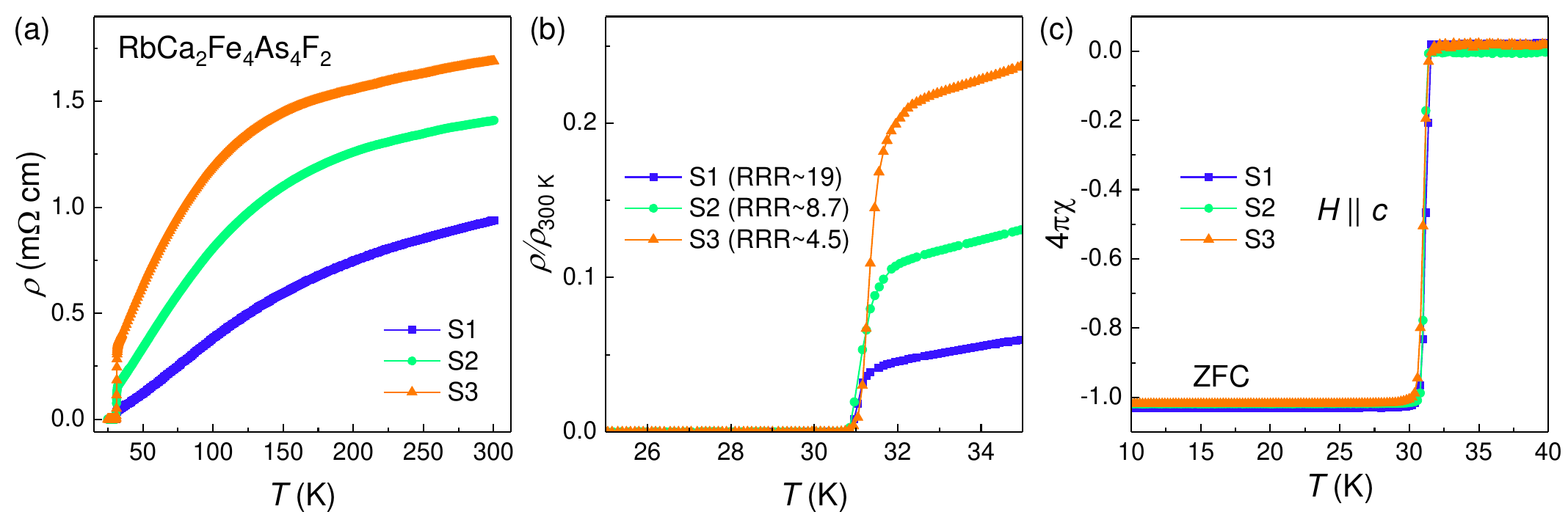}
  \begin{center}
\caption{\label{Fig1}
(a) The temperature dependence of the in-plane resistivity $\rho$(\emph{T}) of three
selected RbCa$_2$Fe$_4$As$_4$F$_2$ single crystals.
(b) Magnification of superconducting transition region normalized to the
resistivity at 300 K.
(c) Temperature dependence of susceptibility under an applied magnetic field of 5
Oe with \emph{H}$\parallel$\emph{c}.
 }
  \end{center}
\end{figure*}

Recently, a new type of stoichiometric iron-based superconductors (IBSs)
\emph{A}Ca$_2$Fe$_4$As$_4$F$_2$ (\emph{A} = K, Rb, Cs), namely, 12442-type was
discovered by the intergrowth of 1111-type CaFeAsF and 122-type
\emph{A}Fe$_2$As$_2$ (\emph{A} = K, Rb, Cs) \cite{755},
consisting of double Fe$_2$As$_2$ layers separated by insulating Ca$_2$F$_2$
layers. Such unique double Fe$_2$As$_2$ layered structure resembles the double
CuO$_2$ layers in cuprate superconductors La$_{2-x}$Sr$_{x}$CaCu$_2$O$_6$ and
Bi$_2$Sr$_2$CaCu$_2$O$_{8+\delta}$ \cite{1933, 2116}.
It manifests a quasi two-dimensional (2D) electronic behavior with a
significant anisotropy of normal-state resistivity
$\rho$$_\textup{c}$/$\rho$$_\textup{ab}$ $>$ 100, and a large upper critical field ($H_\textup{c2}$) anisotropy $\sim$ 8 \cite{1535,
1487}, comparable to those of cuprate superconductors.
The inelastic neutron scattering study has revealed a 2D spin resonant mode with
downward dispersion \cite{1824}, also resembling the behavior observed in cuprates.
Moreover, the pronounced resistive tail of superconducting transition under
magnetic field and relatively low irreversibility field ($H_\textup{irr}$)
indicate that the coupling between double Fe$_2$As$_2$ layers is weaker than most IBSs \cite{1535, 801, 766}.
On the other hand, the SMP effect and the critical current density
$J_\textup{c}$ in the 12442 system are strongly sample-dependent.
For example, the self-field $J_\textup{c}$ of KCa$_2$Fe$_4$As$_4$F$_2$ single
crystal is almost one order of magnitude higher than that of
Rb/CsCa$_2$Fe$_4$As$_4$F$_2$ \cite{1535, 766, 1520},
while a unique SMP effect that shows a non-monotonic variation with temperature is
observed in RbCa$_2$Fe$_4$As$_4$F$_2$ singles crystals \cite{1520}.
To avoid the additional effects introduced by chemical doping,
the stoichiometric \emph{A}Ca$_2$Fe$_4$As$_4$F$_2$ (\emph{A} = K, Rb, Cs) single
crystals with unique intergrowth structure can be regarded as a good candidate to
study the vortex dynamics,
especially the main factors of governing SMP phenomenon.

In this paper, we study the vortex dynamics of RbCa$_2$Fe$_4$As$_4$F$_2$ based on
three selected single crystals with identical superconducting transition temperature $T_\textup{c}$ $\sim$ 31 K but with
different levels of disorder/defect.
It is found that the one with moderate disorder/defects shows a pronounced
non-monotonic SMP effect. Some expanded Ca$_2$F$_2$ layered and dislocation
defects of RbFe$_2$As$_2$ layers are found by microstructure study.
Furthermore, magnetic relaxation measurement reveals that the SMP is strongly associated with the E-P vortex transition.
The systematic evolution of SMP indicates the complex vortex motion of the 12442 system, which
provides insights into the vortex dynamics for novel superconductors with
intergrowth structure.

\section{Experimental Details}
Single crystals of RbCa$_2$Fe$_4$As$_4$F$_2$ were grown by the self-flux method,
details of the crystal growth are given in our previous report \cite{1487}.
X-ray diffraction (XRD) was characterized via a commercial Rigaku diffractometer
with Cu \emph{K}$\alpha$ radiation.
Elemental analysis was performed by a scanning electron microscope equipped with an energy dispersive x-ray spectroscopy probe.
Structure and elemental analysis reveal that only (00\underline{2\emph{l}})
(\emph{l} diffraction peaks are detected and the average atom rations are almost
consistent with the nominal stoichiometry \cite{1487, 1520}.
The in-plane electrical resistivity was carried out by the standard four-probe
method on a Physical Property Measurement System (PPMS, Quantum Design).
Magnetization measurement was preformed by the VSM (vibrating sample magnetometer) option of
PPMS.
The values of Jc are calculated using the Bean mode $J_\textup{c}$ =
20$\Delta$\emph{M}/[\emph{a}(1--\emph{a}/3\emph{b})] \cite{1924},
where $\Delta$\emph{M} = \emph{M}$_\textup{down}$--\emph{M}$_\textup{up}$,
\emph{M}$_\textup{down}$ and \emph{M}$_\textup{up}$ are the magnetization measured
with decreasing and increasing applied field, respectively, \emph{a} and \emph{b}
are sample widths (\emph{a} $<$ \emph{b}).
The normalized magnetic relaxation rate,
\emph{S}(=$|$d$\ln$\emph{M}/dln\emph{t}$|$), was measured by tracing the decay of
magnetization with time \emph{M}(\emph{t}) due to creep motion of vortices for one
hour,
where \emph{t} is the time from the moment when the critical state is prepared.
The cross-sectional observations were investigated by aberration-corrected
high-resolution transmission electron microscopy (TEM, Titan Themis3 G2 300) on
the thin specimen (thickness $\sim$ 50 nm ) prepared by a focused ion beam
instrument (Helios NanoLab G3 UC).

\section{Results and Discussion}

\begin{figure*}
\includegraphics[width=42pc]{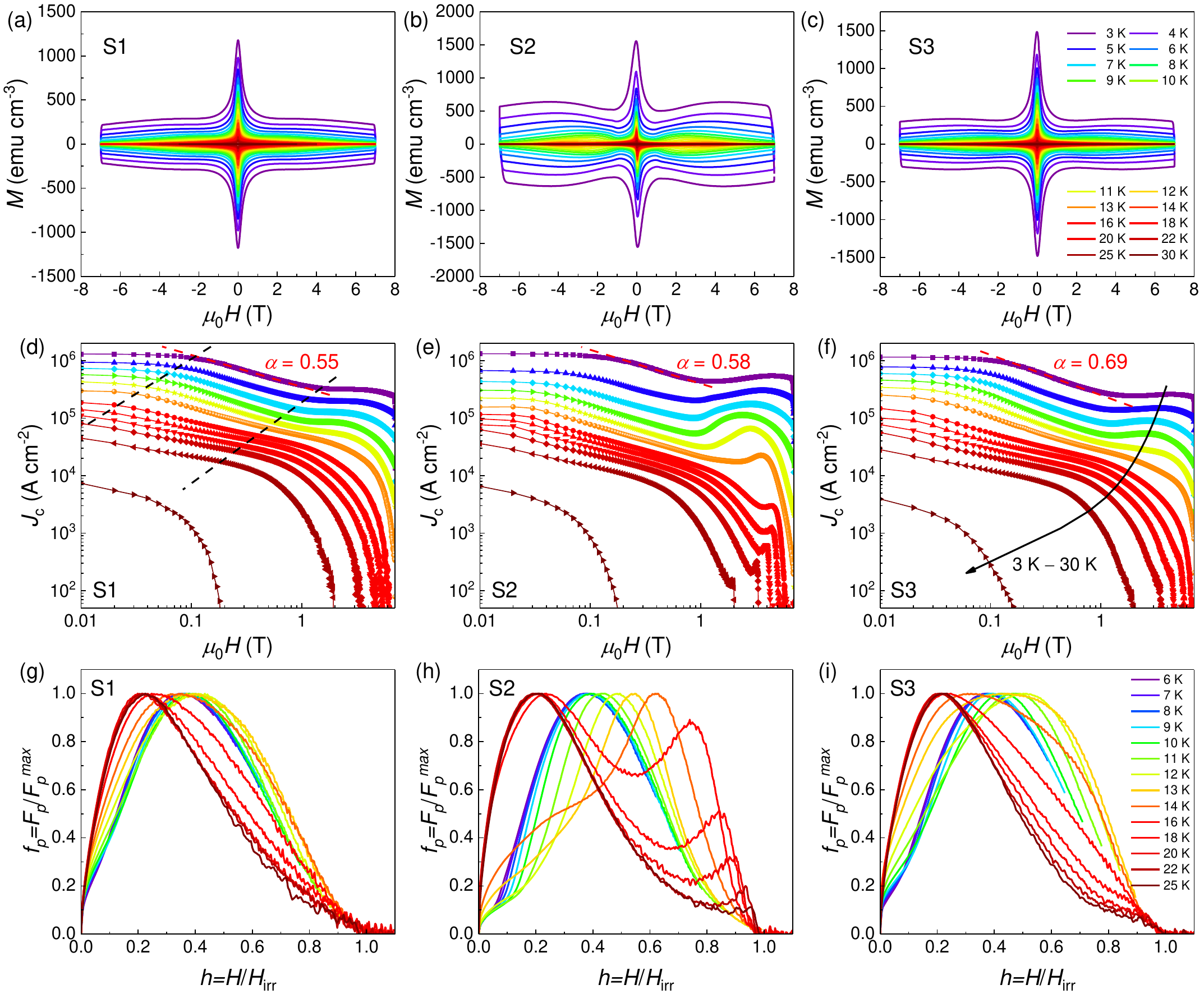}
  \begin{center}
\caption{\label{Fig2}
(a)-(c) Magnetization hysteresis loops at various temperatures ranging from 3 to
30 K for sample S1, S2, and S3, respectively.
(d)-(f) The corresponding field dependence of critical current density at constant
temperatures calculated by using the Bean model.
(g)-(i) The scaling of the normalized pinning force density \emph{f}$_\textup{p}$
= \emph{F}$_\textup{p}$/\emph{F}$_\textup{p}^\textup{max}$ for different
temperatures as a function of the reduced field \emph{h} =
\emph{H}/$H_\textup{irr}$.
}
  \end{center}
\end{figure*}

The in-plane resistivity $\rho$(\emph{T}) curves for
three RbCa$_2$Fe$_4$As$_4$F$_2$ single crystals (named by S1, S2, and S3) are
shown in Fig. \ref{Fig1}(a).
All three samples exhibit proximate metal conductive behavior and
coherent-incoherent transition in the middle temperature region, which is commonly
observed in 12442-type IBSs \cite{755, 761, 758}.
The values of residual resistivity ratio (RRR), characterizing the level of disorder/defect and
defined as $\rho$(300 K)/$\rho$(32.5 K), are estimated to be $\sim$ 19 for sample
S1, $\sim$ 8.7 for sample S2, and $\sim$ 4.5 for sample S3, respectively.
The superconducting transition temperatures $T_\textup{c}$ determined by the zero
resistivity are almost identical with a value of 31 K, regardless of the RRR values.
This indicates that the disorder/defects only destroy local superconductivity and have little effect on $T_\textup{c}$.
Superconductivity was also confirmed by the
susceptibility measurement with zero-field-cooling (ZFC) model, as shown in Fig. \ref{Fig1}(c).
The onsets of diamagnetism are in good agreement with the resistivity data.
The sharp superconducting transition width $\Delta$$T_\textup{c}$, defined
as the temperature difference between 10$\%$ and 90$\%$ of susceptibility, is less than 1 K for all crystals, indicating the homogeneous distribution
of disorder/defects.

To study the effect of disorder/defect on the critical current density, we measured the
magnetization hysteresis loops at various temperatures ranging from 3 to 30 K for
the three selected crystals, as shown in Figs. \ref{Fig2}(a)-(c).
The symmetric loops suggest that the bulk pinning rather than surface or
geometrical barriers is dominant.
What deserves more attention is the pronounced SMP observed in the sample S2,
which is obviously different from the other two crystals
with more or less disorder/defects.
Such difference can also be detected from the field dependence of $J_\textup{c}$
calculated using the Bean model \cite{1924} (see Figs. \ref{Fig2}(d)-(f)).
As seen, the values of the self-field $J_\textup{c}$ for three
crystals are almost the same, slightly larger than 10$^6$ A cm$^{-2}$ at 3 K.
$J_\textup{c}$ changes little at low fields, which can be associated with
single-vortex region.
With increasing magnetic field, $J_\textup{c}$ follows a power-law behavior
$J_\textup{c}$ $\propto$ $H^{\alpha}$ with $\alpha$ = 0.55 for sample S1, 0.58 for
sample S2, and 0.69 for sample S3, respectively.
Such power-law behavior is also observed in most IBSs, which can be attributed to
the sparse strong point pinning by sparse nano-sized defects \cite{1916, 1917}.
In general, due to the addition of random point defects by chemical doping or irradiation, the level of disorder/defect in the nanoscale increases, and the $\alpha$ value decreases gradually,
as revealed in H$^{+}$-irradiated GdBa$_2$Cu$_3$O$_{7-\delta}$ and Ni doped CaKFe$_4$As$_4$ \cite{1925, 1312}.
Therefore, the different values of $\alpha$ indicate the different levels of defect in three crystals.

\begin{figure}
\includegraphics[width=21pc]{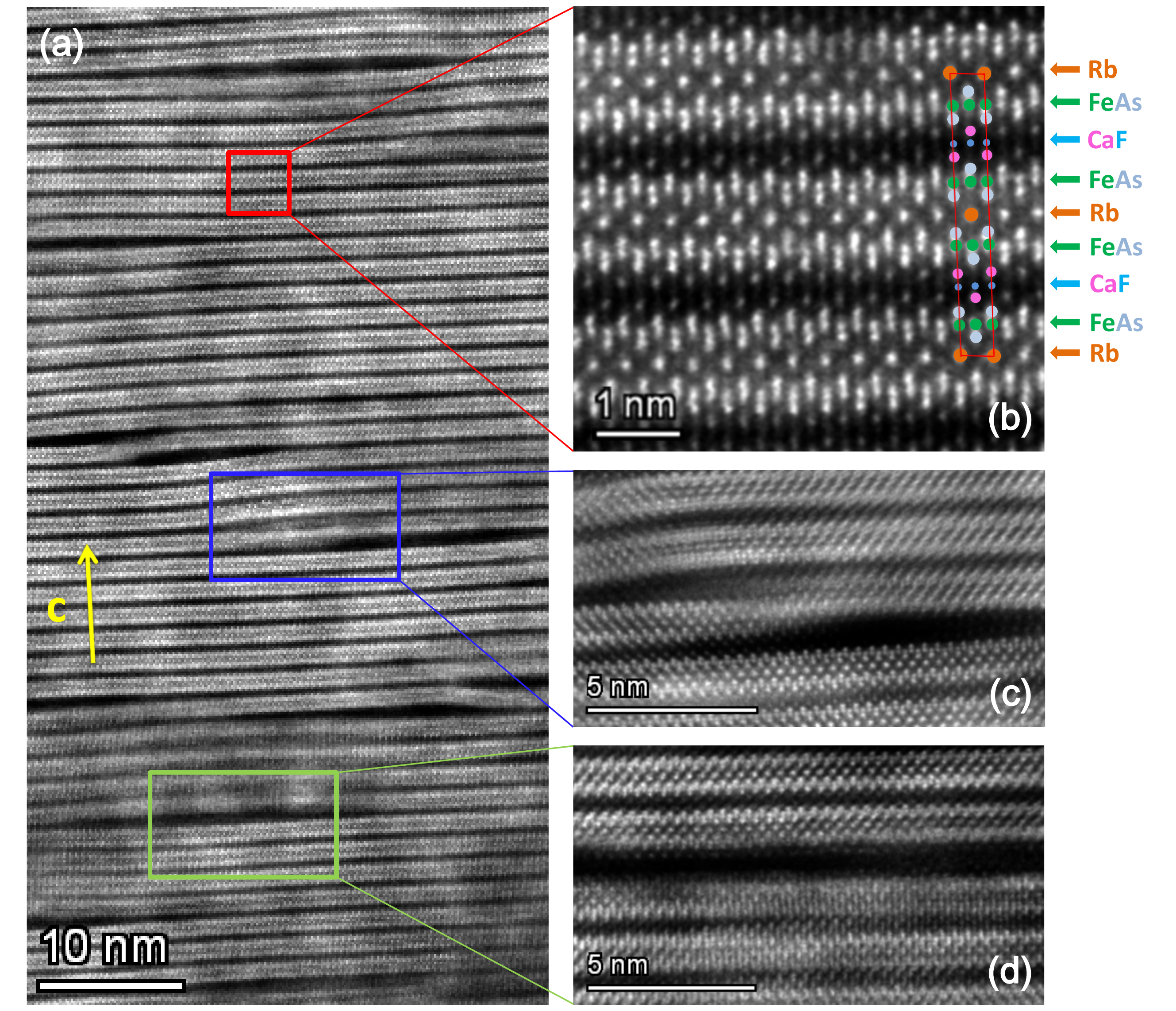}
  \begin{center}
\caption{\label{Fig3}
(a) The microstructures of RbCa$_2$Fe$_4$As$_4$F$_2$ single crystal S2
investigated by STEM.
(b) High resolution STEM image around regular lattice structure, and atomic
arrangements are demonstrated.
(c) A type of dislocation defects observed in RbFe$_2$As$_2$ layer, and stretched
lattice in the upper right corner.
(d) Another type of expanded Ca$_2$F$_2$ layered defects. }
  \end{center}
\end{figure}

Depending on the magnetic field strength, the behavior of $J_\textup{c}$(\emph{H})
with SMP is generally divided into several regimes:
(I) a low-field regime associated with the single-vortex regime;
(II) a power law dependence $J_\textup{c}$ $\propto$ $H^{\alpha}$ related to
strong pinning centers;
(III) a SMP regime related to random disorder and usually related to a crossover
from E-P relaxation of vortex lattice;
and (IV) a high-field regime characterized by a fast drop in
$J_\textup{c}$(\emph{H}) \cite{347}.
The magnetic field $H_\textup{sp}$ at which the SMP appears usually decreases with
increasing temperature in most of the type-II superconductors.
Noticeably, $H_\textup{sp}$ of sample S2 decreases at first, then increases with
increasing temperature, and finally decreases with a steep transition adjacent to
$H_\textup{irr}$ \cite{1520},
which can be seen more clearly in Fig. \ref{Fig5}.
However, the other two samples present unobvious or faded SMP characteristics at low temperatures.
Therefore, it is plausible to expect that the SMP effect in
RbCa$_2$Fe$_4$As$_4$F$_2$ is closely related to the level of disorder/defect.

To gain more insight into the underlying pinning mechanisms,
the pinning force (\emph{F}$_\textup{p}$ =
$\mu$$_0$\emph{H}$\times$$J_\textup{c}$) is calculated.
As proposed by Dew-Hughes \cite{1918}, the pinning mechanism does not change with
temperature if the normalized pinning force \emph{f}$_\textup{p}$ =
$F_\textup{p}$/$F_\textup{p}^\textup{max}$ at different temperatures
can be scaled into a unique curve by the reduced field \emph{h} =
\emph{H}/$H_\textup{c2}$, where \emph{F}$_\textup{p}^\textup{max}$ is the maximum
pinning force, and $H_\textup{c2}$ is the upper critical field.
However, the upper critical field is generally too high to be obtained in
high-temperature superconductors, so the irreversibility field $H_\textup{irr}$
is widely used in cuprates and IBSs \cite{389, 341, 363}.
The deviation from the scaling usually suggests the change in vortex-lattice
period or the various size of pinning centers. In Figs. \ref{Fig2}(g)-(i), we
present the \emph{f}$_\textup{p}$ vs \emph{h} = \emph{H}/$H_\textup{irr}$ for three crystals.
$H_\textup{irr}$ is determined from $J_\textup{c}$(\emph{H}) curves with the
criterion of \emph{J}$_\textup{c}$ = 30 A cm$^{-2}$ at high temperatures, or
extrapolating to \emph{J}$_\textup{c}$ = 0 in the \emph{J}$_\textup{c}^{1/2}$ vs
\emph{H} at low temperatures.
Except from the SMP in S2, \emph{f}$_\textup{p}$(\emph{h}) shows similar behavior in the three crystals.
At low temperatures, the peaks of \emph{f}$_\textup{p}$(\emph{h}) are located at \emph{h}$_\textup{max}$ $\sim$ 0.4-0.5.
Similar results have been observed in other superconductors such as Ba$_{0.68}$K$_{0.32}$Fe$_2$As$_2$ (\emph{h}$_\textup{max}$ $\sim$ 0.43)
\cite{451}, BaFe$_{1.9}$Ni$_{0.1}$As$_2$ (\emph{h}$_\textup{max}$ $\sim$ 0.4)
\cite{1736, 367}, and Ba$_{0.66}$K$_{0.32}$BiO$_{3+\delta}$
(\emph{h}$_\textup{max}$ $\sim$ 0.47) \cite{1831}.
According to Dew-Hughes model \cite{1918}, the present case of
\emph{h}$_\textup{max}$ $<$ 0.5 is suggestive of $\delta$\emph{l}-type pinning,
which arises from a spatial variation in the mean free path of charge carriers,
and the pinning is due to the presence of a large density of point-like defect
centers whose dimensions are smaller than the intervortex distance.
Besides, the peaks in the low field region are located at \emph{h}$_\textup{max}$ $\sim$ 0.2, which is the characteristic of surface strong pinning, such as the planar defects.
Obviously, the characteristic of two \emph{f}$_\textup{p}$(\emph{h}) peaks is more obvious in sample S2 with the SMP.
Therefore, the vortex pinning in RbCa$_2$Fe$_4$As$_4$F$_2$ at low temperatures
mainly comes from the weak collective pinning at high magnetic field,
while it is mainly dominated by the strong pinning at high temperatures.

Because of the non-monotonic temperature dependence of $H_\textup{sp}$ at \emph{T}
$>$ 9 K in sample S2 \cite{1520}, the \emph{f}$_\textup{p}$(\emph{h}) curves
affected by SMP diverge gradually and cannot be scaled.
When it reaches the maximum point of $H_\textup{sp}$ near 16 K, the \emph{f}$_\textup{p}$(\emph{h}) curve shows two peaks in the low and high field regions, respectively.
In addition to scaling peak at low \emph{h}, the \emph{f}$_\textup{p}$(\emph{h}) peak induced by SMP
effect at high temperatures moves gradually to $H_\textup{irr}$, and reduces
quickly with the decrease of \emph{f}$_\textup{p}$.
The step-like transition approaching to $H_\textup{irr}$ was also reported in NbSe$_2$
\cite{1881}, optimally doped Ba$_{1-x}$K$_x$BiO$_3$ \cite{1108}, and untwinned
YBa$_2$Cu$_3$O$_y$ \cite{1864},
and it is generally understood as the vortex melting transition.

In order to understand the origin of vortex pinning and the type of defect  in RbCa$_2$Fe$_4$As$_4$F$_2$,
the cross-sectional microstructures of the typical sample S2 were
explored by high resolution TEM measurement.
A typical layered structure with light and dark stripes is observed, as shown in
Fig. \ref{Fig3}(a).
The white stripes in Fig. \ref{Fig3}(b) correspond to the Rb and Fe$_2$As$_2$
layers,
while the dark stripes correspond to the Ca$_2$F$_2$ layers. The complete lattice
structure and atomic arrangement of RbCa$_2$Fe$_4$As$_4$F$_2$ are demonstrated by
the colored spheres.
Microstructural investigations in stoichiometric 1144-type CaKFe$_4$As$_4$
superconductors have revealed defects of fine-sized stacking fault of
CaFe$_2$As$_2$ and/or KFe$_2$As$_2$ layers as well as the lattice mismatch stress
inside the grains \cite{1280, 1611, 1577}.
However, neither linearly grown single layers nor step-like stacking fault layers
can be visible evidently in RbCa$_2$Fe$_4$As$_4$F$_2$.
It also seems reasonable that the inter and intra double Fe$_2$As$_2$ layers have
different symmetries. It is impossible for 1111- and 122-type structures to form
stacking faults in the same layer or ladder-like structure.
But it is interesting to find two other types of unique defects:
(i) large number of obviously expanded Ca$_2$F$_2$ layers with various width shown
in Fig. \ref{Fig3}(d),
and (ii) obvious dislocations existing in the same layer of RbFe$_2$As$_2$ shown
in Fig. \ref{Fig3}(c).
The torsional dislocation in the RbFe$_2$As$_2$ layer seems to be accompanied by
the stretched Ca$_2$F$_2$ layers on both sides.
All these exhibit significant lattice stress \cite{1535}, which may be the main
reason for the irregular light and dark change in Fig. \ref{Fig3}(a) and the
obvious stretched lattice in the upper right corner of Fig. \ref{Fig3}(c).
It has been reported that there is weak anisotropic $J_\textup{c}$ in 12442 system
with \emph{J}$_\textup{c}^{H||ab}$ $<$ \emph{J}$_\textup{c}^{H||c}$ \cite{1535,
766},
which is obviously different from 1144 system with significantly anisotropic
$J_\textup{c}$ and $J_\textup{c}^{H||ab}$ $>$ $J_\textup{c}^{H||c}$ \cite{1280,
1611, 1449, 1281}.
In CaKFe$_4$As$_4$, fine-sized planar defects of CaFe$_2$As$_2$ and/or
KFe$_2$As$_2$ layers along \emph{ab} plane act as pinning centers for vortices
when \emph{H}${||}$\emph{ab}, and significantly increase $J_\textup{c}$.
While for RbCa$_2$Fe$_4$As$_4$F$_2$ single crystals, appropriate amounts of
lamellar defects with smaller width and thinner thickness,
the lattice mismatch stress, and chemical inhomogeneity, work together as pinning
centers in both directions,
showing relatively smaller critical current anisotropy and novel SMP phenomenon.

\begin{figure}
\includegraphics[width=21pc]{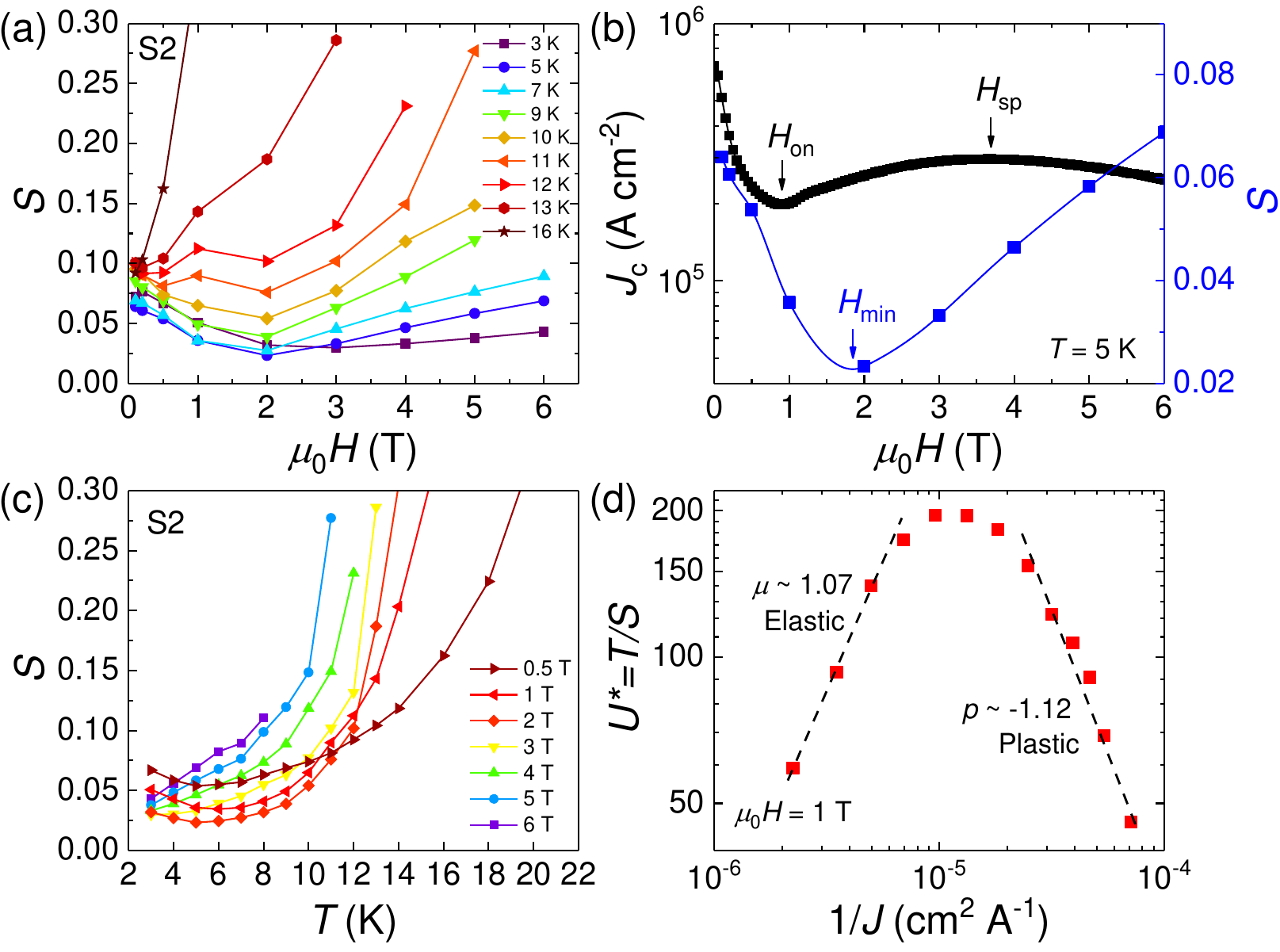}
  \begin{center}
\caption{\label{Fig4}
Measurement of magnetic relaxation of sample S2.
(a) The magnetic field dependence of the normalized relaxation rate \emph{S} under
different temperatures.
(b) Field dependence of critical current density $J_\textup{c}$ and relaxation
rate \emph{S} at \emph{T} = 5 K.
(c) The temperature dependence of the normalized relaxation rate \emph{S} under
different magnetic fields.
(d) Inverse current density dependence of effective pinning energy \emph{U}$^*$ at
$\mu$$_0$\emph{H} = 1 T.}
  \end{center}
\end{figure}

\begin{figure*}
\includegraphics[width=40pc]{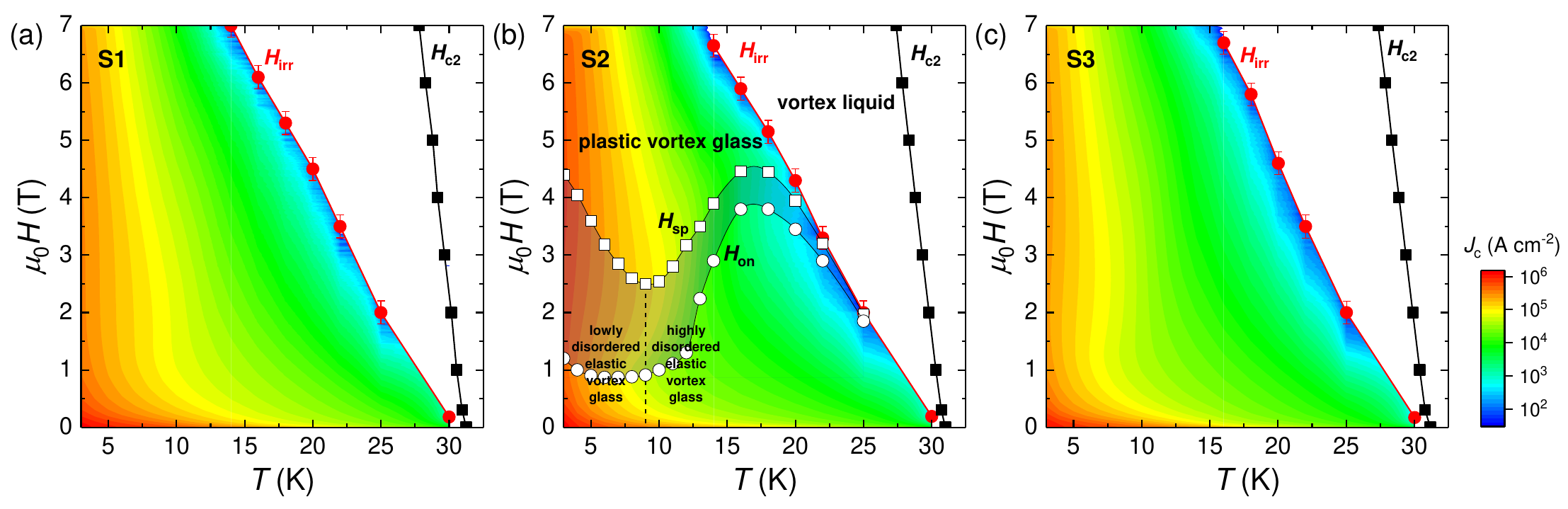}
  \begin{center}
\caption{\label{Fig5}
Vortex phase diagrams of three selected RbCa$_2$Fe$_4$As$_4$F$_2$ single crystals
(a) S1, (b) S2, and (c) S3 with different disorder/defect levels, respectively.
}
  \end{center}
\end{figure*}

It has been proposed that the SMP effect is generally present in samples with
moderate pinning strength,
possibly associated with a crossover from E-P transition
\cite{347, 1312}.
To further verify and explore the origin of the SMP in the high magnetic field region, measurement of the normalized magnetic relaxation rate,
\emph{S}(=$|$d$\ln$\emph{M}/dln\emph{t}$|$),
is performed as a function of temperature for several applied magnetic
fields on sample S2. Fig. \ref{Fig4}(a) shows the magnetic field dependence of the
normalized relaxation rate.
With increasing magnetic field, \emph{S} at low temperatures first drops to the value
of $\sim$ 0.03, followed by a gradual increasing,
which is attributed to successive change of vortex bundle size from a single
vortex to small bundle and large bundle regimes \cite{1857}.
At high temperatures, \emph{S} increases rapidly.
The values are much larger than those of conventional superconductors, but
comparable to those 11-type \cite{841}, 122-type \cite{347}, and 1144-type IBSs
\cite{1312}, exhibiting a universal "giant flux creep".
It has been observed that the minimum in the \emph{S}(\emph{H}) curves is generally accompanied with the presence of SMP phenomenon \cite{342, 346, 423, 421}.
In order to probe the correlation between the creep rate and SMP,
\emph{S}(\emph{H}) curves at selective temperatures are also plotted together
with the $J_\textup{c}$(\emph{H}) curve at the same temperatures.
As shown in Fig. \ref{Fig4}(b), the minimum $H_\textup{min}$ in the
\emph{S}(\emph{H}) curve is located between the onset ($H_\textup{on}$) of SMP and
the second peak position ($H_\textup{sp}$).
This feature has also been found in FeSe$_{0.5}$Te$_{0.5}$ \cite{1843},
Ba(Fe$_{0.93}$Co$_{0.07}$)$_2$As$_2$ \cite{347},
Ca$_{0.8}$La$_{0.2}$Fe$_{1-x}$Co$_x$As$_2$ \cite{363}, and many superconductors
with SMP, in agreement with the picture of E-P vortex transition.
Fig. \ref{Fig4}(c) shows the temperature dependence of magnetic relaxation rate
\emph{S} under different magnetic fields.
A plateau in the intermediate-temperature range with a high vortex creep rate
\emph{S} $\sim$ 0.05 is detected, as previously observed in
YBa$_2$Cu$_3$O$_{7-\delta}$ \cite{428}, SmFeAsO$_{0.9}$F$_{0.1}$ \cite{431},
Ba(Fe$_{1-x}$Co$_x$)$_2$As$_2$ \cite{343},
Ca$_{0.8}$La$_{0.2}$Fe$_{1-x}$Co$_x$As$_2$ \cite{363}, and FeSe$_{0.6}$Te$_{0.4}$
\cite{529}, which can be interpreted by the collective creep theory \cite{428}.

It is well known that the IBSs usually exhibit a giant flux creep rate, and it can be
well described by the collective pinning theory,
which is characterized by the current density \emph{J} dependence of the
activation energy \emph{U}.
The effective activation energy as a function of current density \emph{J} is given
by interpolation formula \emph{U}(\emph{J}) =
(\emph{U}$_0$/$\mu$)[(\emph{J}$_\textup{c0}$/\emph{J})$^{\mu}$--1], where
\emph{U}$_0$ is the collective pinning barrier in the absence of flux creep,
\emph{J}$_\textup{c0}$ is the temperature-dependent critical current density in
the absence of flux creep, and ${\mu}$ is the vortex pinning regime-dependent
glassy exponent \cite{1857}.
In this theory, the glassy exponent ${\mu}$ is related to the vortex-bundle size,
and it is predicted as ${\mu}$ = 1/7, 3/2 or 5/2, 7/9 for single-vortex,
small-bundle, and large-bundle regimes, respectively \cite{1857, 366}.
By defining the effective pinning barrier \emph{U}$^*$ = \emph{T}/\emph{S} and
combining the interpolation formula, \emph{U}$^*$ can be calculated as:
\emph{U}$^*$ = \emph{U}$_0$+$\mu$\emph{T}ln(\emph{t}/\emph{t$_0$}) =
\emph{U}$_0$(\emph{J}$_\textup{c0}$/\emph{J})$^{\mu}$.
Thus, the slope in the double logarithmic plot of \emph{U}$^*$ vs 1/\emph{J}
represents the value of ${\mu}$, as manifested in Fig. \ref{Fig4}(d).
To avoid the effect of fast relaxation at high temperatures and high magnetic
fields, $\mu$ $\sim$ 1.07 was evaluated under 1 T in low temperature region.
This value is the prediction of single-vortex and small-bundle regimes, indicating
the contributions from different pinning types.
On the other hand, the evaluated negative slopes \emph{p} $\sim$ --1.12 are also
observed at small \emph{J} region, which is often denoted as plastic creep
scenario with \emph{p} $\sim$ --0.5 \cite{1905}.
It should be pointed out that E-P vortex phase transition is a necessary
condition for the occurrence of SMP, but not a sufficient factor.
For example, it exists in FeSe single crystals without SMP \cite{841}, but is absent in the different Co doped CaFe$_2$As$_2$ whose vortex dynamics are plastic creeping rather than collective creep model \cite{337}.
In addition, for the Co doped BaFe$_2$As$_2$ in the very underdoped and overdoped region,
the SMP effect becomes invisible or very weak \cite{343}.
Therefore, the existence of the E-P crossover as well as the moderate disorder can lead to the emergence of a typical SMP.
In other words, the combination of sparse strong pinning centers and dense weak
pinning centers is beneficial to SMP.

From the above experimental results, we depict the vortex phase diagrams of three
RbCa$_2$Fe$_4$As$_4$F$_2$ single crystals shown in Figs. \ref{Fig5}(a)-(c).
The color contour represents the critical current density $J_\textup{c}$. The
orange and yellow range with $J_\textup{c}$ $>$ 10$^5$ A cm$^{-2}$ extends to a
high magnetic field region, indicating its excellent current-carrying capacity.
$H_\textup{irr}$ is the irreversibility field, and the upper critical field
$H_\textup{c2}$ is determined by the resistive measurement with criterions of 90\%
of the normal state resistivity $\rho$$_\textup{n}$ under magnetic field.
Above $H_\textup{c2}$(\emph{T}) line, the system enters into the normal state. Below it,
it changes into vortex liquid state, and a unique vortex slush phase exists near
the vortex glass temperature \cite{1520}.
In the vortex phase diagram of sample S2,
$H_\textup{sp}$(\emph{T}) divides the critical current region into two parts
beneath $H_\textup{irr}$(\emph{T}).
Below $H_\textup{sp}$(\emph{T}), the vortex motion can be well interpreted with
the collective vortex motion.
With increasing magnetic field, the flux creep becomes faster, and the system
enters into the plastic creep regime.
The elastic vortex glass region can be divided into the lowly and highly
disordered vortex glass state by abnormal point around 9 K (represented by the
dotted line) \cite{1520}.
In fact, the actual E-P transition occurred at the characteristic field
$H_\textup{min}$ between $H_\textup{sp}$ and $H_\textup{on}$ \cite{363}.

For the non-monotonic behavior of $H_\textup{sp}$, it is found that the width of
the SMP as marked by black shadow decreases gradually with increasing temperature.
At low temperatures, E-P phase transition has been confirmed by the
results of normalized magnetic relaxation.
The dense weak pinning centers dominate the collective pinning. The pinning
potential increases first and then decreases with magnetic field.
In the high temperature region, strong pinning centers are dominant, and the
pinning potential decreases monotonously.
At the same time, it should be accompanied by the change of magnetic relaxation mode.
In the region of low temperature and low field, since the elastic creep of flux bundle is dominant, there is less disorder and stronger rigidity.
On the other hand, in the high temperature and low field region, the flux softens gradually.
It changes to the plastic creep of single vortex with increasing disorder, and is characterized by the increase of creep.
Therefore, the SMP in the high temperature region is still dominated by E-P phase transition,
and the width of SMP decreases monotonously with increasing temperature.
When approaching the $H_\textup{irr}$(\emph{T}) line, the vortex phase transition from elastic to plastic is very close to the first-order flux melting transition in low temperature superconductors in the high field region.
The SMP gradually evolves into a step-like transition and then becomes a peak shape.
We suspect that the step-like behavior may be a crossover from the second-order (E-P transition) to first-order transition (vortex melting transition).
Certainly, this calls for further investigations both experimentally and theoretically \cite{1881}.
It is noted that the E-P phase transition reflected by this non-monotonic
SMP in the 12442 system just connects the SMP of high temperature superconductors
at low fields and the SMP of low temperature superconductors at high fields.

\section{Conclusion}

In summary, we have systematically studied three RbCa$_2$Fe$_4$As$_4$F$_2$
single crystals that have almost the same $T_\textup{c}$ $\sim$ 31 K but
with different amounts of disorder/defects.
Only the sample S2 with moderate disorder/defects shows significant SMP effect,
although their self-field $J_\textup{c}$ are basically similar.
This result manifests the importance of a certain amount of disorder/defects in the formation of SMP.
The evolution of the normalized pinning force density
demonstrates the type of pinning changing from the dominant density weak pinning
at low temperatures to strong pinning at high temperatures.
Some expanded Ca$_2$F$_2$ layers and dislocation defects in RbFe$_2$As$_2$ layers
may be the main cause of non-monotonic SMP phenomenon.
The step-like transition of anomalous SMP connects the SMP of high temperature
superconductors and that of low temperature superconductors, revealing the
possible universal origin related to the E-P phase transition.

\begin{acknowledgments}
This work was supported by the Strategic Priority Research Program (B) of the
Chinese Academy of Sciences (Grant No. XDB25000000), the National Key R\&D Program
of China (Grant No. 2018YFA0704300), the National Natural
Science Foundation of China (Grant Nos. 12204265 and 12204487), and the Fundamental Research Funds for the Central Universities.
\end{acknowledgments}

Xiaolei Yi and Xiangzhuo Xing contributed equally to this work.

\bibliography{references}


\end{document}